\documentclass[reprint,nofootinbib,superscriptaddress,amsmath,amssymb,aps,prl]{revtex4-2}
\usepackage{amsmath,mathrsfs,amsbsy,color,graphicx,bm,amsthm,amsfonts}
\usepackage{bbm}
\usepackage{times}
\usepackage{units}
\usepackage{dcolumn}
\usepackage{graphicx}
\usepackage{epsfig}
\usepackage{epstopdf}
\usepackage[colorlinks,linkcolor=red,anchorcolor=green,citecolor=blue,CJKbookmarks=True]{hyperref}
\DeclareMathSymbol{\shortminus}{\mathbin}{AMSa}{"39}
\usepackage{mathrsfs}
\usepackage{braket}
\usepackage{amssymb}
\usepackage{txfonts}
\usepackage{float}
\usepackage{enumitem}
\usepackage{multirow}
\usepackage{orcidlink}

\newcommand{\meq}[1]{(\ref{#1})}

\begin{document}
\title{Macroscopic Optical Nonreciprocity: A Black Hole as an Optical Diode}

\author{Wentao Liu\orcidlink{0009-0008-9257-8155}}
\affiliation{Lanzhou Center for Theoretical Physics, Key Laboratory of Theoretical Physics of Gansu Province, 
Key Laboratory of Quantum Theory and Applications of MoE,
Gansu Provincial Research Center for Basic Disciplines of Quantum Physics, 
Lanzhou University, Lanzhou 730000, China}
\affiliation{Institute of Theoretical Physics $\&$ Research Center of Gravitation,
Lanzhou University, Lanzhou 730000, China}

\author{Di Wu\orcidlink{0000-0002-2509-6729}}
\affiliation{School of Physics and Astronomy, China West Normal University, Nanchong, Sichuan 637002, P. R. China}

\author{Xiongjun Fang\orcidlink{0000-0002-4270-3332}}
\affiliation{Department of Physics, Key Laboratory of Low Dimensional Quantum Structures and Quantum Control of Ministry of Education, and Synergetic Innovation Center for Quantum Effects and Applications, Hunan Normal University, Changsha, Hunan 410081, P. R. China}

\author{Yu-Xiao Liu\orcidlink{0000-0002-4117-4176}}
\email[]{liuyx@lzu.edu.cn (Corresponding authors)} 
\affiliation{Lanzhou Center for Theoretical Physics, Key Laboratory of Theoretical Physics of Gansu Province, 
Key Laboratory of Quantum Theory and Applications of MoE, Gansu Provincial Research Center for Basic Disciplines of Quantum Physics, 
Lanzhou University, Lanzhou 730000, China}
\affiliation{Institute of Theoretical Physics $\&$ Research Center of Gravitation,
Lanzhou University, Lanzhou 730000, China}

\author{Jieci Wang\orcidlink{0000-0001-5072-3096}}
\email[]{jcwang@hunnu.edu.cn (Corresponding authors)} 
\affiliation{Department of Physics, Key Laboratory of Low Dimensional Quantum Structures and Quantum Control of Ministry of Education, and Synergetic Innovation Center for Quantum Effects and Applications, Hunan Normal University, Changsha, Hunan 410081, P. R. China}

\begin{abstract}



Optical reciprocity—the principle that light retraces the same path when source and detector are interchanged—is a foundational concept in geometric optics. 
In this Letter, we demonstrate that this “symmetry-protected” behavior can be qualitatively overturned in a rotating black hole when spontaneous Lorentz symmetry breaking introduces a nonminimally coupled background structure with a preferred direction.   
Through numerical ray-tracing simulations, we reveal a striking macroscopic signature: upon optical-path reversal achieved by exchanging the source and the observer, the shadow of the same black hole morphs from a quasi-symmetric rugby-ball shape into a distinct teardrop profile. 
This high-contrast nonreciprocity effectively turns the black hole into a cosmic-scale optical diode, offering a novel pathway to probe fundamental symmetries using current and next-generation horizon-scale imaging.


\end{abstract}

\vspace*{0.2cm}
\maketitle

\begin{figure*}[t]
\centering
\includegraphics[width=0.95\linewidth]{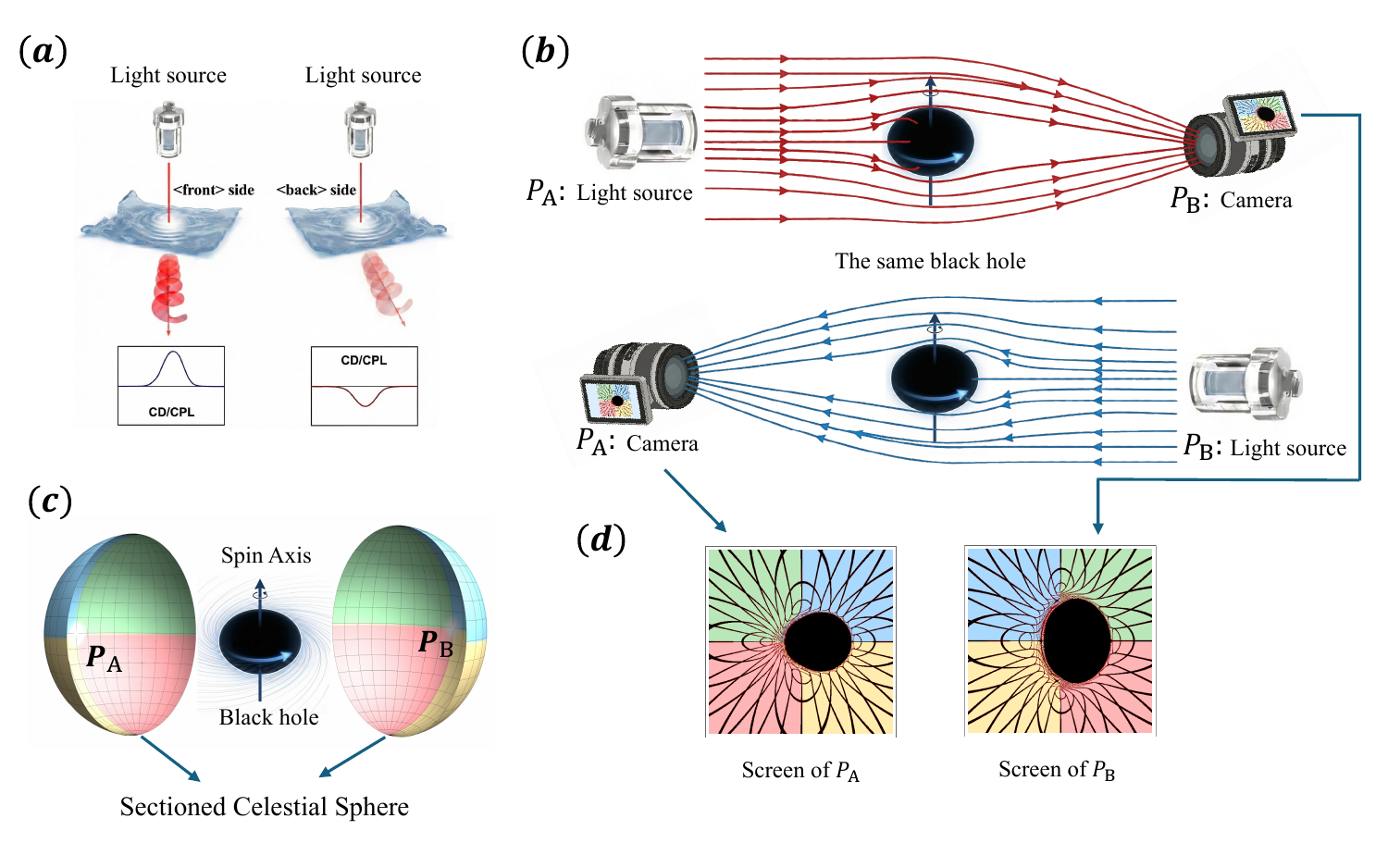}
\caption{{Rotating black hole as a macroscopic optical diode.} 
(a) {Nonreciprocal optical systems:} The system exhibits distinct optical responses when excited from opposite directions. 
(b) {Optical path reversal:} Implemented by exchanging the source ($P_\text{A}$) and the observer ($P_\text{B}$). The LSB-spin coupling breaks the symmetry between forward and backward paths. 
(c) {Simulation configuration:} Rotating black hole at the center of a sectioned celestial sphere with spin along the $+z$ direction. 
(d) {Nonreciprocal shadow profiles:} Upon reversing the light path, the shadow morphs from a rugby-ball shape (right) to a teardrop profile (left), manifesting high-contrast macroscopic optical nonreciprocity.
}
\label{fig1}
\end{figure*}

Optical reciprocity, requiring identical paths for counterpropagating light, is a cornerstone of geometric optics in static spacetimes of General Relativity (GR).  
In stationary spacetimes such as the Kerr black hole, this reciprocity is indeed broken by frame-dragging; however, the violation is strictly constrained by the discrete $t\!-\!\varphi$ symmetry of the spacetime geometry \cite{Carter:1968rr}.
Hence, optical nonreciprocity in GR can be understood as a ``symmetry‑protected" reciprocity: a rigid, purely geometric effect of rotation that does not support the tunable, diode‑like nonreciprocal dynamics known in other contexts.
In contrast, within the domain of cavity quantum electrodynamics and nanophotonics, optical nonreciprocity, i.e., the breakdown of reciprocity under the exchange of source and detector, has been actively explored and experimentally realized. 
A paradigmatic approach involves the optical Sagnac–Fizeau effect in spinning resonators, where controlled mechanical rotation explicitly breaks time-reversal symmetry to induce high-contrast optical isolation \cite{maayani2018flying}.
These studies have demonstrated high-contrast nonreciprocal light transmission at the level of linear optical response.
Further theoretical work has revealed unconventional and nonreciprocal photon blockade at the quantum statistical level, leading to asymmetric optical responses \cite{Huang:2018mjv}.
From a phenomenological perspective, optical nonreciprocity can be most clearly defined by comparing the optical response under the exchange of illumination and detection directions, while keeping the system itself unchanged. 
As illustrated in Fig. $\!$\ref{fig1}(a), when the same optical excitation is applied from opposite sides of the system, the resulting circular dichroism (CD) or circularly polarized luminescence (CPL) signals exhibit qualitatively different spectral features upon reversal of the propagation direction, which directly demonstrates the breakdown of optical reciprocity.

However, the symmetry-protected reciprocity of light propagation in GR is a consequence of the spacetime symmetries underlying the Lorentz-invariant geometric structure.
Proposals from quantum gravity posit that Lorentz symmetry may be spontaneously broken at high energies, potentially challenging these fundamental invariances  \cite{Carroll:2001ws}.
In this Letter, we demonstrate that spontaneous Lorentz symmetry breaking (LSB) transforms a rotating black hole into a macroscopic ``gravitational optical diode'' that exhibits genuine optical nonreciprocity.
Specifically, LSB induces an effective refractive index for photon propagation that depends on the orientation of the photon momentum relative to the black hole spin.
Within this effective geometric optics description, the rotating black hole can be viewed as a macroscopic optical element, whose nonreciprocal behavior can be qualitatively understood by analogy with a nonreciprocal photonic system.
Guided by this analogy, we perform numerical ray-tracing simulations in which the rotating black hole acts as an effective optical element illuminated by a distant background source, with incident radiation propagating perpendicular to the spin axis. 
Photons escaping to infinity illuminate the observation screen, while captured photons define the black hole shadow.
In standard GR, optical reciprocity implies that the shadow profile remains geometrically invariant under the exchange of the source and the observer.
In contrast, in the presence of a Lorentz-violating vector field with nonminimal coupling, this symmetry-protected reciprocity is explicitly broken. 
As shown in Fig. $\!$\ref{fig1}(b)-(d), the shadow morphology exhibits a striking directional dependence: the profile resembles a quasi-symmetric rugby-ball shape in one configuration, but transforms into a distinct teardrop geometry upon reversing the light path.
Given the idealized nature of the numerical setup, this effect represents macroscopic optical nonreciprocity induced by LSB, which breaks the visual symmetry of the black hole shadow between opposite observation directions.
In this context, the black hole functions as a cosmic-scale optical isolator, regulating photon-capture rates based on the propagation direction—a behavior strictly forbidden in standard GR.

To elucidate the physical origin of this macroscopic high-contrast nonreciprocity, we now turn to the underlying theoretical framework governed by a Lorentz-violating vector field.
Because the associated symmetry breaking induces inherently nonreciprocal optical responses, its effects are expected to become observationally accessible only in extreme gravitational settings.
As prime examples of such environments and the focal point where strong gravity and quantum effects potentially intersect, black holes serve as ideal testbeds for constraining deviations from GR.
Our analysis is based on the Einstein-bumblebee gravity model  \cite{Kostelecky:2003fs}, where a dynamical vector field $B_\mu$, which is non-minimally coupled to gravity, acquires a non-vanishing vacuum expectation value (VEV), resulting in spontaneous LSB.
This model provides a concrete framework that extends the standard formalism of GR and is particularly suited for our purposes.
To capture the non-perturbative effects of Lorentz violation in rotating spacetimes, we need to start from the pioneering work of Poulis et al., which provides the exact axisymmetric black hole solutions.
The modified spacetime geometry is described by the exact solution \cite{Poulis:2021nqh}:
\begin{equation}\label{ds2}
ds^2 = ds^2_\text{Kerr} + \frac{\ell}{\Delta}\left(r dr \pm a\sqrt{\Delta}\cos\theta d\theta\right)^2.
\end{equation}
Here, $ds^2_\text{Kerr}$ denotes the Kerr line element, and $\ell$ is the Lorentz-violation parameter associated with the VEV of the bumblebee field $B_\mu$ and quantifies deviations from the standard Kerr geometry.
Crucially, the expansion of the quadratic term generates a non-diagonal metric component $g_{r\theta} \propto \pm \ell a \cos\theta$. 
This term strictly couples the radial and latitudinal motions, breaking the reflection symmetry of photon geodesics. 
Note that $\Delta=r^2-2Mr+a^2$ retains its Kerr form, preserving the event horizon structure.
To probe the observational signatures of this symmetry breaking, we analyze the photon trajectories governed by the Hamiltonian constraint:
\begin{equation}
\mathcal{H} = \frac{1}{2}g^{\mu\nu}p_{\mu}p_{\nu} = 0.
\end{equation}
The nondiagonal metric term physically introduces a momentum cross-term $g^{r\theta}p_r p_\theta$ in the Hamiltonian. 
Unlike the Kerr spacetime, where the separation of variables is guaranteed by the Carter constant, this cross term renders the dynamical system non-integrable. 
Consequently, we employ ray-tracing to numerically integrate these coupled equations \cite{Cunha:2015yba,Zhong:2021mty,Bacchini:2021fig,Liu:2025wwq}. 
Our simulations reveal that the chiral asymmetry induced by the non-diagonal term $g^{r\theta}$ is indeed the physical origin of the macroscopic teardrop shadows and optical nonreciprocity shown in Fig. \!\ref{fig1}(d).
Crucially, this phenomenon stems from the interplay between the VEV of the $B_\mu$ field and the black hole spin. 
In the limit of vanishing VEV $\ell \!\to\! 0$, the preferred spacetime direction disappears, restoring the standard D-shaped Kerr shadow. 
Similarly, in the static limit $a \!\to\! 0$, the critical cross-term $g^{r\theta}$ vanishes, thereby recovering optical reciprocity.

Next, we detail the simulation configuration illustrated in Fig. \!\ref{fig1}(b)-(d) to numerically implement the theoretical proposal of using the black hole as an optical diode.
Given that the exact rotating solution with the non-minimally coupled $B_\mu$ field admits two distinct branches, we choose, for simplicity, to focus on the case corresponding to the $ (+) $ sign of the $ (\pm) $ term in metric \meq{ds2}.
We place the black hole at the center of a large celestial sphere, with its spin angular momentum aligned with the positive $z-$axis, as illustrated in Fig. \!\ref{fig1}(c).
Note that the black hole size in the figure is exaggerated for visualization; in the actual simulations, the radius of the celestial sphere is much larger than the black hole scale.
We define two antipodal points, $P_\text{A}$ and $P_\text{B}$, located on the equator of the celestial sphere.
These points alternate roles: when one serves as the light source, the other acts as the observer.
The distinct colors on the celestial sphere (Fig. \ref{fig1}(c)) encode the directional information of incident light.
With the black hole present, optical reciprocity ensures a one-to-one mapping of the light paths between $P_\text{A}$ and $P_\text{B}$.
In the presence of the black hole, however, some photons are captured by the event horizon, creating a shadow.
The remaining photons, carrying the angular directional information from the celestial sphere, reach the observer and form the distorted four-quadrant patterns shown in Fig. \ref{fig1}(d).
We first consider the configuration in the upper panel of Fig. \!\ref{fig1}(b), where $P_\text{A}$ acts as the source and $P_\text{B}$ as the observer.
Light rays traversing the foreground (facing the reader) correspond to prograde orbits.
Due to frame-dragging, these photons have a higher escape probability and readily reach $P_\text{B}$.
In contrast, rays traversing the background correspond to retrograde orbits.
These photons encounter a larger effective capture cross-section and are more likely to be trapped by the event horizon.
For the light path from $P_\text{A}$ to $P_\text{B}$, modulated by the LSB field, this asymmetry manifests as a rugby-ball-shaped shadow in the right panel of Fig. \ref{fig1}(d), with a contour that extends prominently towards the background.
Upon reversing the light path, for the same black hole, we focus on the configuration in the lower panel of Fig. \ref{fig1}(b), with $P_\text{A}$ now acting as the observer.
In this case, the shadow bulges towards the foreground, yielding the peculiar teardrop shape shown in the left panel of Fig. \!\ref{fig1}(d).
The violation arises from spacetime anisotropy defined by the privileged direction, breaking the symmetry-protected reciprocity and eliminating the geometric symmetry between forward and backward paths.

Functionally, this directional dependence mirrors that of a nonreciprocal optical diode. 
While standard geometric asymmetry permits reciprocal transmission, the LSB interaction strictly breaks the symmetry between forward and backward paths.
In our gravitational context, the event horizon acts as a spin-dependent filter: the distinct capture cross-sections for forward ($P_\text{A} \!\to \! P_\text{B}$) and backward ($P_\text{B}\!\to\! P_\text{A}$) propagation imply that the black hole possesses direction-dependent transmissivity.
Consequently, activated by the black hole spin\footnote{The black hole spin acts as the effective ``control knob'' of the optical response. In the static limit  \cite{Casana2018}, the LSB field decouples analytically from photon trajectories, such that the geodesic equations reduce to the Schwarzschild form, restoring optical reciprocity and rendering the LSB sector optically invisible.}, the spontaneous LSB effectively induces a macroscopic rectification of the photon flux.

\begin{figure}[h]
\centering
\includegraphics[width=0.95\linewidth]{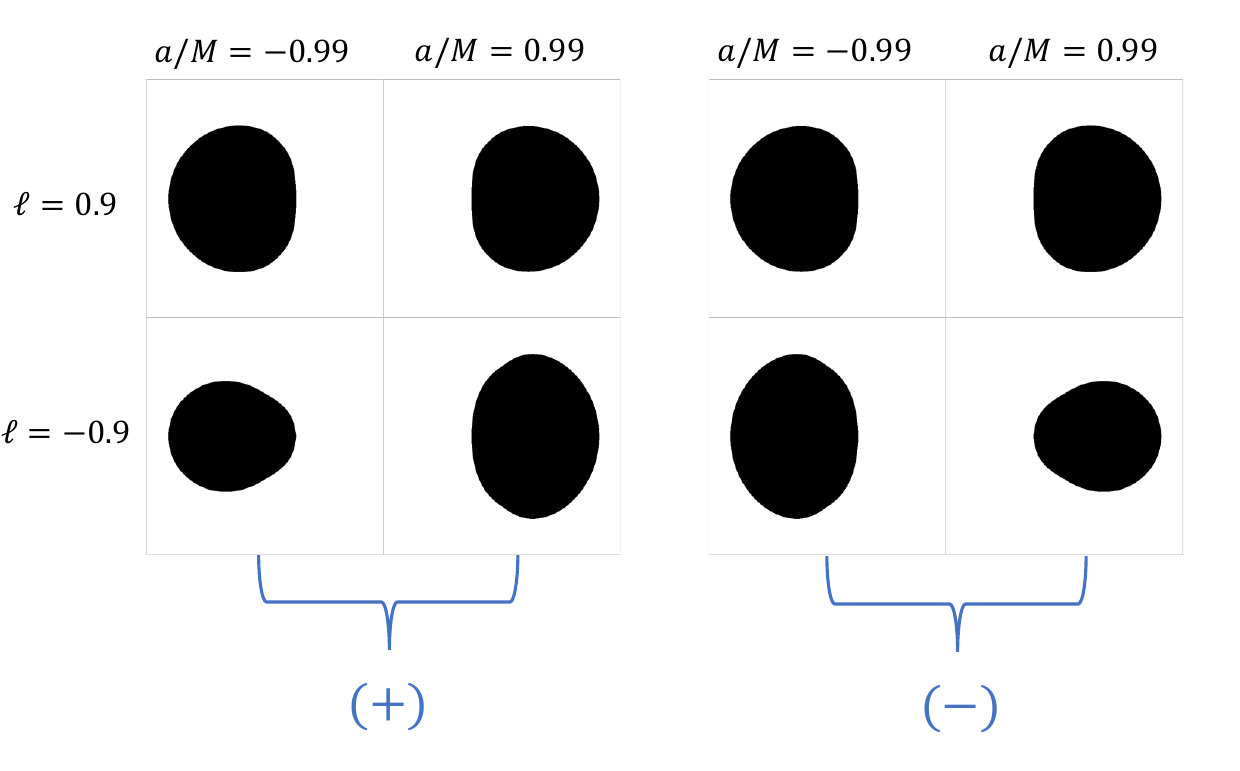}
\caption{Rotating Lorentz-violating black holes manifest as ``conjugate'' black holes in their shadow contours.
The (+) black hole is on the left, and the (-) black hole is on the right, both under near extreme conditions.
Note that the large parameter $\ell\!=\! \pm0.9$ is chosen to exaggerate the topological deformation for visual clarity; the qualitative nonreciprocity persists for any non-vanishing Lorentz violation.}
\label{fig2}
\end{figure}

\begin{figure*}[t]
\centering
\includegraphics[width=0.98\linewidth]{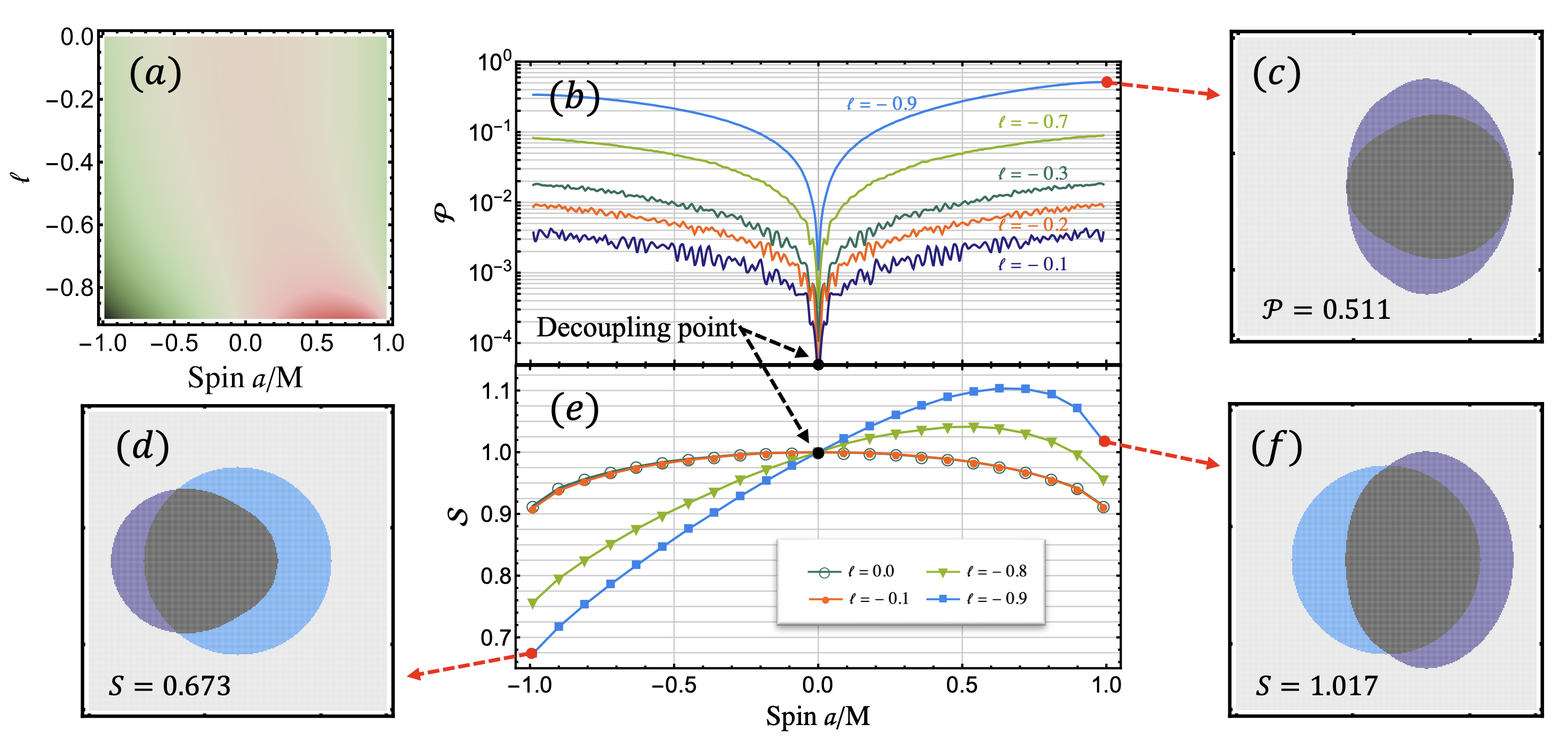}
\caption{Quantitative analysis of the macroscopic optical nonreciprocity. 
(a) Phase diagram showing how the normalized area $\mathcal{S}$ deviates from unity. 
(b) and (e) show the variations of the nonreciprocity index $\mathcal{P}$ and $\mathcal{S}$, respectively. 
(d) and (f) visualize the physical definition of $\mathcal{S}$ by highlighting the capture-area deviation from the Schwarzschild baseline. 
(c) elucidates the underlying principle of $\mathcal{P}$ by superimposing forward and backward profiles to reveal the directional area asymmetry.}
\label{fig3}
\end{figure*}
So, a natural question then arises as to why such high-contrast nonreciprocity emerges.
In fact, the symmetry associated with the exchange of light paths in spacetime geometry still exists, but it manifests as a generalized symmetry relating the two branches of the bumblebee gravity solution.
We can note that metric \meq{ds2} describes two black hole spacetimes, differentiated by the sign in front of the cross term $ g_{r\theta} $, which can be either $ (+) $ or $ (-) $. 
While standard uniqueness theorems typically dictate a single physical manifold \cite{Chrusciel:2012jk}, this solution intrinsically emerges as a 	`twin' structure.
Crucially, exchanging the source and the observer is mathematically equivalent to reversing the black hole spin.
As shown in Fig. \ref{fig2}, the $(+)$ black hole is not self-symmetric under spin reversal; rather, it is symmetric only with respect to the counter-rotating black hole in the $(-)$ branch.
In terms of shadow morphology, rotating black holes with the $ (+) $ and $ (-) $ signatures thus appear conjugate. 
Interestingly, this confirms that it is sufficient to restrict our analysis to the $(+)$ metric, as adopted in the earlier setup of this Letter.

To quantify the macroscopic nonreciprocal effect shown in Fig. \ref{fig2}, we adopt the shadow area as a geometric observable, since in the geometric-optics limit it is determined by the photon capture region and therefore provides a direct proxy for the absorbed luminous flux.
We then introduce two dimensionless quantities,
\begin{equation}
\mathcal{S}=A_\text{Lv}^{(a,\ell)}/A_\text{Sch}^{(0,0)},\quad\quad
\mathcal{P}=\bigg|A_\text{Lv}^{(a,\ell)}/A_\text{Lv}^{(-a,\ell)}-1\bigg|,
\end{equation}
where $A^{(a,\ell)}$ denotes the shadow-area fraction, defined as the fraction of screen pixels belonging to rays captured by the black hole \cite{Liu:2025wwq}.
Here, the normalized area $\mathcal{S}$ measures the overall change in the effective photon-capture area relative to the Schwarzschild case, while the nonreciprocity index $\mathcal{P}$ quantifies the asymmetry between the forward and backward configurations and thus provides a direct measure of optical nonreciprocity.

We then perform a systematic scan over the black hole parameter space $(a/M\in[-0,99,0.99],\ell\in[-0.9,0])$ at a resolution of $256\times256$, and the resulting phase diagram is shown in Fig.\,\ref{fig3}(a). 
The dark red region in the lower right and the dark green region in the lower left correspond to the parameter domains where the positive and negative deviations of $\mathcal{S}$ from unity are most pronounced, respectively. 
Their representative visual comparisons with the Schwarzschild shadow are displayed in Fig. \ref{fig3}(f) and Fig. \ref{fig3}(d). 
Here, the blue and purple pixels denote the shadow regions of the Schwarzschild and Lorentz-violating black holes, respectively, while the overlap is shown in black. 
The non-overlapping regions therefore directly visualize the area excess or deficit induced by LSB.
Figure \ref{fig3}(e) presents representative data corresponding to Fig. \!\ref{fig3}(a). 
For relatively large LSB, the nonreciprocal effect is clearly enhanced as the black hole spin increases. 
Even for small LSB, a slight asymmetry remains visible: compared with the Kerr case $\ell\!=\!0$, the $\ell\!=\!-0.1$ case already exhibits a discernible deviation from reciprocity. 
Figure \ref{fig3}(b) makes this even clearer: except at the decoupling point, $\mathcal{P}\!\neq\!0$ throughout the parameter space, indicating that optical nonreciprocity is a generic feature of rotating black holes in this theory.
We also show in Fig. \ref{fig3}(c) a representative case where the nonreciprocity index reaches its maximum. 
For a more direct comparison, we mirror-flip one of the shadows and superpose it onto the other. 
It is then clear that, for the same parameters, the teardrop-shaped shadow is enclosed by the olive-shaped one, while their endpoints along the symmetry axis nearly coincide. 
This indicates that the former always has a smaller area than the latter. 
The non-overlapping region therefore provides a direct visualization of the capture-area asymmetry between the two propagation directions, consistent with the behavior of $\mathcal{P}$ shown in Fig. \ref{fig3}(b).
In addition, because our results are extracted from ray-tracing simulations, they inevitably contain finite-resolution numerical artifacts associated with pixelization; these effects can be systematically suppressed as the resolution is increased.
While existing weak-field tests place stringent constraints on LSB, we adopt \(|\ell|\sim\mathcal{O}(0.1)\) in our visualizations only to amplify the shadow deformation; the underlying nonreciprocity persists even for much smaller Lorentz violation.

In this Letter, we demonstrate that spontaneous LSB can transform a rotating black hole into a macroscopic ``gravitationaloptical diode" that exhibits genuine optical nonreciprocity. 
This phenomenon stems from the nonminimal coupling, which endows the spacetime with a privileged direction defined by the VEV of the bumblebee field, effectively establishing a preferred axis for photon propagation \cite{Ovcharenko:2026rvj}.
The resulting anisotropy, together with  the black hole spin, induces a nondiagonal metric component $g_{r\theta}$--absent in standard GR--that fundamentally alters photon kinematics by introducing a parity-violating cross-term $p_r p_\theta$ in the geodesic Hamiltonian.
Consequently, photons propagating parallel or antiparallel to this axis experience distinct effective refractive indices, leading to a transition from a quasi-symmetric rugby-ball shape to a distorted teardrop geometry upon optical path reversal.
From an observational standpoint, these distinctive shadow signatures offer a promising avenue for testing fundamental symmetries with the Event Horizon Telescope (EHT) \cite{EventHorizonTelescope:2019dse,EventHorizonTelescope:2022wkp}. 
While current EHT observations of M87* and Sgr A* are broadly consistent with GR, their finite resolution leaves open the possibility of exotic physics.
The next-generation EHT  \cite{Johnson:2023ynn}, with its substantially enhanced angular resolution and dynamic range, could make the macroscopic high‑contrast nonreciprocal effect predicted here detectable.  In particular, the directional dependence of the shadow morphology may become observable.
Identifying such a violation of the reflection symmetry in the shadow profile would not only constrain the Lorentz-violating parameter space but also provide  a ``smoking gun" signature of physics beyond the Standard Model in the strong-gravity regime.
Moreover, the two branches of rotating Lorentz-violating black holes exhibit a conjugate relationship, as reflected in their shadows.
Given that the bumblebee gravity model emerges as a low-energy limit of string theory \cite{Kostelecky:1988zi,Bluhm:2004ep}, this naturally raises a compelling conjecture: could the discrete conjugate symmetry observed here be a four-dimensional manifestation of dual black hole pairs existing in higher-dimensional spacetimes?



\textit{Acknowledgments.}\textemdash
This work is supported in part by the National Natural Science Foundation of China (Grants No. 12475056, No. 12374408, No. 12247101, No. 12547147), Gansu Province’s Top Leading Talent Support Plan, the Natural Science Foundation of Gansu Province (No. 22JR5RA389), and the 111 Project (Grant No. B20063), the China Postdoctoral Science Foundation (Grant No. 2025M783393), the Sichuan Science and Technology Program under Grant No. 2026NSFSC0021.

\appendix

%

\end{document}